\documentclass[aps,pra,twocolumn %,superscriptaddress
]{revtex4-1}
\usepackage{graphicx}
\usepackage{amsfonts}
\usepackage{amssymb}
\usepackage{array}
\usepackage{amsmath}
\usepackage{verbatim} 
\usepackage{hyperref}
\usepackage{color}
\usepackage{epstopdf}
\usepackage{bbold}
\usepackage{longtable}

\newcommand{\ket}[1]{\left\vert#1\right\rangle}
\newcommand{\bra}[1]{\left\langle#1\right\vert}

\def\bra#1{\langle #1|}
\def\ket#1{|#1 \rangle}

\begin{document}
\title{Generation of hybrid entanglement between 
a single-photon polarization qubit and a coherent state}
\author{Hyukjoon Kwon and Hyunseok Jeong}
%\email{h.jeong37@gmail.com}
\affiliation{Center for Macroscopic Quantum Control, Department of Physics and Astronomy, Seoul National University, Seoul, 151-742, Korea}
\date{\today}
\begin{abstract}
We propose a scheme to generate entanglement between a single-photon qubit in the polarization basis
and a coherent state of light. 
The required resources are
a superposition of coherent states,
a polarization entangled photon pair, beam splitters, the displacement operation, and four photodetectors. 
Even when realistic detectors with a limited efficiency are used,
an arbitrarily high fidelity can be obtained by adjusting a beam-splitter ratio and the displacement amplitude
at the price of reducing the success probability. 
Our analysis shows that high fidelities may be obtained using on-off detectors with low efficiencies and available resource states under current technology.
\end{abstract}
\maketitle

\section{Introduction}
Entangled light fields have been extensively explored as tools for testing quantum mechanics and resources for quantum information processing. 
An intriguing challenge in this subject is to entangle different types of states of light such as microscopic and macroscopic states  or wavelike and particlelike states 
\cite{DeMartini1998,DeMartini2008,Sekatski2010,Sekatski2012,Ghobadi2013,Bruno2013,
Lvovsky2013,Andersen2013,Jeong2014,Morin2014}.
Some of those states have been found useful for quantum information applications 
%\cite{Rigas2006,Wittmann2010, Spagnolo2011,Stobinska2011, Park2012,Kwon2013,SWLee2013,Stobinska2014}.
\cite{Park2012,Kwon2013,SWLee2013,Stobinska2014}.
Recently, hybrid entanglement between a single photon in the polarization basis and a coherent state was found to be particularly useful for  loophole-free Bell inequality tests \cite{Kwon2013}, deterministic quantum teleportation, and resource-efficient quantum computation \cite{SWLee2013}.
It was also shown that this type of hybrid entanglement can be purified using linear optical elements and the parity check gates \cite{Sheng2013}.
While single photons are regarded  as  nonclassical states as light quanta,
coherent states are considered to be classical states as their $P$ functions are well defined \cite{Mandel1986}
and they are robust against decoherence as ``pointer states'' \cite{Zurek1993}.
In this regard, the hybrid entanglement is closely related to Schr\"odinger's {\it Gedankenexperiment},
where the fate of a classical object, the cat, is entangled with the state of a single atom
\cite{Schrodinger1935}.

Very recently, approximate implementations of hybrid entanglement 
between a qubit of the vacuum and single photon and a qubit of coherent states
were demonstrated using the photon addition and subtraction techniques \cite{Jeong2014,Morin2014}.
The state explored in Ref.~\cite{Jeong2014} was in the form of $\ket{0}\ket{\alpha} + \ket{1}\ket{-\alpha}$ while 
a similar state of $(\ket{0}+\ket{1})\ket{\alpha} + (\ket{0}-\ket{1})\ket{-\alpha}$ was
approximately demonstrated in Ref.~\cite{Morin2014},
where $|0\rangle$ is the vacuum, $|1\rangle$ is the single photon, and $|\pm \alpha\rangle$ are coherent states of amplitudes $ \pm \alpha$.
However, the state required to perform the aforementioned applications in Refs.~\cite{Kwon2013, SWLee2013} was in fact in the form of $\ket{H}\ket{\alpha} + \ket{V}\ket{-\alpha}$; i.e., the first mode should be in a {\it definite  single-photon state} in the horizontal ($H$) or vertical ($V$) polarization. This type of hybrid entanglement, despite its usefulness, cannot be generated using the photon addition or subtraction as performed in Refs.~\cite{Jeong2014, Morin2014} because the first mode should be in a single photon state with definitely one photon.
In principle, a cross-Kerr nonlinear interaction can be used to generate the required form of hybrid entanglement \cite{Gerry1999,Jeong2005}, but it is a highly demanding task to achieve a clean nonlinear interaction using current technology \cite{Shapiro2006, Shapiro2007, Banacloche2010,He2012}.
% Cite Shapiro etc
 
In this article, we suggest a nondeterministic scheme  to generate the desired form of hybrid entanglement  between a single-photon polarization qubit and a coherent-state field. 
Our scheme requires 
a superposition of coherent states (SCS), $\ket{\alpha} + \ket{-\alpha}$
\cite{Ourjoumtsev2006,Neergaard-Nielsen2006,Ourjoumtsev2007, Takahashi2008, Gerritt2010}, and 
a polarization entangled photon pair, $\ket{H}\ket{V}+ \ket{V}\ket{H}$, as resources, in addition to beam splitters, the displacement operation and four  photodetectors. 
We find that even when inefficient detectors  are used, an arbitrarily high fidelity can be obtained by adjusting a beam-splitter ratio, and the displacement amplitude.
Our proposal is experimentally feasible using a squeezed single photon (or a squeezed vacuum state) as a good approximation of an ideal SCS \cite{Lund2004}. Remarkably, reasonably high fidelities may still be obtained using on-off detectors with low efficiencies and available resource states under current technology.

\section{Generation Scheme}
%{\it Generation scheme.---}

We aim to generate the optical hybrid state
\begin{equation}
\ket{\Psi_\varphi(\alpha_f)}_{AB} = \frac{1}{\sqrt{2}} \left( \ket{H}_A \ket{\alpha_f}_B +e^{i\varphi} \ket{V}_A \ket{-\alpha_f}_B \right),
\label{HybridState}
\end{equation}
where $|\pm \alpha_f\rangle_B$ are coherent states in the field mode $B$ and $\varphi$ is a relative phase factor.
As discussed in Ref.~\cite{Review2014}, this type of state shows obvious properties as macroscopic entanglement when $\alpha$ is sufficiently large.
For example, it is straightforward to show that the measure $\cal I$ as a macroscopic superposition \cite{CWLee2011} for this state has its maximum value ${\cal I}=\alpha_f^2+1$,  i.e., the average photon number of the state.
A classification of hybrid entanglement was attempted \cite{class}, according to which
the state in Eq.~(\ref{HybridState}) is categorized as a discrete-variable-like hybrid entanglement.
This type of entanglement was also characterized by a matrix Wigner function in the context of trapped ions \cite{Vogel1997}.

\begin{figure}[t]
\includegraphics[width=0.8\linewidth]{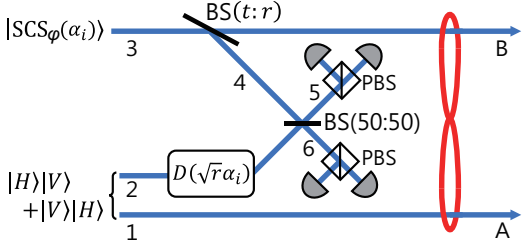}
\caption{(Color online) Generation scheme for hybrid entanglement. The beam-splitter reflectivity $r$ and the amplitude $\alpha_i$ of the SCS determine the amplitude $\sqrt{r}\alpha_i$ of the displacement operation.  }
\label{GenerationScheme}
\end{figure}

In order to generate the hybrid entanglement, as shown in Fig.~\ref{GenerationScheme}, we first need to prepare a polarization entangled photon pair and a SCS as
\begin{equation}
 \ket{\chi}_{12} \otimes \ket{{\rm SCS_\varphi}(\alpha_i)}_3,
\end{equation}
where $ \ket{\chi}_{12} = (\ket{H}_1\ket{V}_2 + \ket{V}_1\ket{H}_2)/{\sqrt{2}} $ and
$\ket{{\rm SCS}_\varphi(\alpha_i)}_3 = N_\varphi ( \ket{\alpha_i}_3 + e^{i \varphi} \ket{-\alpha_i}_3)$ with $N_\varphi = (2 + 2 e^{-2 |\alpha_i|^2}\cos{\varphi} )^{-1/2}$.
We suppose that $\alpha_i$ and $\alpha_f$ are real without losing generality throughout the article.
A beam splitter of transmissivity $t$ (reflectivity $r = 1-t$) splits a coherent state $|\alpha\rangle$ into $\ket{\sqrt{r}\alpha}\ket{\sqrt{t}\alpha}$. The unbalanced beam splitter in Fig.~\ref{GenerationScheme} thus transforms   $\ket{{\rm SCS_\varphi}(\alpha_i)}_3$ into
 $\ket{\sqrt{r}\alpha_i}_4\ket{\sqrt{t}\alpha_i}_B + e^{i \varphi} \ket{-\sqrt{r}\alpha_i}_4\ket{-\sqrt{t}\alpha_i}_B$.
At the same time, 
the displacement operation is performed on mode 2 as
$D_2(\sqrt{r} \alpha_i) \left( \ket{H}_A \ket{V}_2 + \ket{V}_A \ket{H}_2 \right)$,
where $D(\alpha)= e^{\alpha \hat{a}^\dagger - \alpha^* \hat{a}} $, and $\hat{a}^\dagger$ and
$\hat{a}$ are the creation and annihilation operators.
The state after the beam splitter and the displacement operation can be expressed as

\begin{equation}
\begin{aligned}
&[D_4(\sqrt{r} \alpha_i) D_B(\sqrt{t} \alpha_i) + e^{i \varphi} D_4(-\sqrt{r} \alpha_i) D_B(-\sqrt{t} \alpha_i)] 
\\
&~~\otimes
 D_2(\sqrt{r} \alpha_i) (\hat{a}^\dagger_{1H}\hat{a}^\dagger_{2V} + \hat{a}^\dagger_{1V} \hat{a}^\dagger_{2H}) 
\ket{0}_1\ket{0}_2\ket{0}_4\ket{0}_B,
\label{eq:step1}
\end{aligned}
\end{equation}
in terms of operators acting on the vacuum states.

A 50:50 beam splitter as shown in Fig.~\ref{GenerationScheme} is then used to mix the reflected part of $\ket{{\rm SCS_\varphi}(\alpha_i)}_3$ (mode 4) and the displaced part of $\ket{\chi}_{12}$ (mode 2) in order to erase `which path' information.
The unitary matrix corresponding to the beam splitter can be represented as

\begin{equation}
\left(
\begin{array}{c}
\hat{a}_6	\\
\hat{a}_5
\end{array}
\right)
=
\left(
\begin{array}{cc}
\cos{\xi}	& -i e^{i \phi} \sin{\xi} \\
-i e^{- i \phi} \sin{\xi} 		& \cos{\xi}
\end{array}\
\right)
\left(
\begin{array}{c}
\hat{a}_4	\\
\hat{a}_2
\end{array}
\right),
\end{equation}
where we choose $\xi=\pi/4$ and $\phi = \pi/2$ to model the 50:50 beam splitter.
The operators of modes 2 and 4 are then transformed  as $\hat{a}_2 \rightarrow (\hat{a}_5 + \hat{a}_6)/\sqrt{2}$ and $\hat{a}_4 \rightarrow (-\hat{a}_5 + \hat{a}_6)/\sqrt{2}$, respectively, and it is also straightforward to show 
$D_4(\alpha) D_2(\beta)\rightarrow D_5[ (-\alpha+\beta)/{\sqrt{2}} ] D_6 [(\alpha+\beta)\sqrt{2}]$
After passing through a 50:50 beam splitter, the operators for modes 2 and 4 evolve as
\begin{equation}
D_4(\alpha)
 D_2(\beta) \hat{a}^\dagger_{2\lambda} 
\rightarrow
 D_5\left( \frac{-\alpha+\beta}{\sqrt{2}} \right) D_6 \left( \frac{\alpha+\beta}{\sqrt{2}} \right)
\frac{\hat{a}^\dagger_{5\lambda} + \hat{a}^\dagger_{6\lambda}}{\sqrt{2}},
\label{eq:step2}
\end{equation}
where $\lambda$ indicates the polarization direction, $H$ or $V$.
By taking $\alpha = \pm \sqrt{r} \alpha_i$ and $\beta = \sqrt{r} \alpha_i$, only one of the displacement operators survives with amplitude $\sqrt{2r}\alpha_i$ in modes 5 and 6, while operators in the other modes, $\hat{a}_A$ and $\hat{a}_B$, remain the same.
Using Eqs.~(\ref{eq:step1}) and (\ref{eq:step2}), we find the state right before reaching the polarizing beam splitters (PBSs)
in Fig.~\ref{GenerationScheme}  as 
\begin{widetext}
\begin{equation}
\begin{aligned}
\ket{\psi_\varphi}=
& \frac{N_\varphi}{2}
\left[ \ket{H}_A {D}_6 (\sqrt{2r}\alpha_i) \left( \ket{V}_5\ket{0}_6 + \ket{0}_5\ket{V}_6 \right) \ket{\sqrt{t} \alpha_i}_B \right. 
 + e^{i \varphi} \left. \ket{H}_A {D}_5 (\sqrt{2r}\alpha_i) \left( \ket{V}_5\ket{0}_6 + \ket{0}_5 \ket{V}_6 \right) \ket{-\sqrt{t} \alpha_i}_B \right. \\
& + \left. \ket{V}_A {D}_6 (\sqrt{2r}\alpha_i) \left( \ket{H}_5\ket{0}_6 + \ket{0}_5\ket{H}_6 \right) \ket{\sqrt{t} \alpha_i}_B \right. 
 + e^{i \varphi} \left. \ket{V}_A {D}_5 (\sqrt{2r}\alpha_i) \left( \ket{H}_5\ket{0}_6 + \ket{0}_5 \ket{H}_6 \right) \ket{-\sqrt{t} \alpha_i}_B \right].
\label{pre-state}
\end{aligned}
\end{equation}
\end{widetext}

The final step is to measure two single photons, one for mode 5 and the other for mode 6, in different polarizations.
The first measurement operator can be expressed as 
\begin{equation}
\Pi = \mathbb{1}_A \otimes \ket{0}\bra{0}_{5H} \otimes \ket{1}\bra{1}_{5V} \otimes \ket{1}\bra{1}_{6H} \otimes \ket{0}\bra{0}_{6V} \otimes \mathbb{1}_B.
\end{equation}
The second and third terms of Eq.~(\ref{pre-state}) are excluded by the conditioning measurement $\Pi$.
It produces to the ideal hybrid state as
\begin{equation}
\rho = \frac{{\rm Tr}_{56} \Big[ {\Pi \ket{\psi_\varphi}\bra{\psi_\varphi}} \Big]}{ \bra{\psi_\varphi} \Pi \ket{\psi_\varphi}} =  \ket{\Psi_\varphi(\alpha_f)}\bra{\Psi_\varphi(\alpha_f)}_{AB}
\end{equation}
where $\alpha_f= \sqrt{t} \alpha_i$.
The success probability to obtain the hybrid state is
\begin{equation}
\begin{aligned}
P^\varphi &= \bra{\psi_\varphi} \Pi \ket{\psi_\varphi}	\\
&= N_\varphi^2 (1-t) \alpha_i^2 e^{-2 (1-t) \alpha_i^2 }
			= N_\varphi^2 (\frac{1}{t}-1) \alpha_f^2 e^{-2 (\frac{1}{t}-1) \alpha_f^2 }.
\end{aligned}
\end{equation}
The success probability  for a given value of $\alpha_i$
can be maximized by taking
$ t= 1- {1}/({2\alpha_i^2})$ with the hybrid state size $\alpha_f = \sqrt{\alpha_i^2 - {1}/{2}}$.
In this case, $P^\varphi$ approaches $1/(8e) \approx 4.60\%$ when the initial amplitude $\alpha_i$ is large enough.

The other measurement event of $\Pi' = \mathbb{1}_A \otimes \ket{1}\bra{1}_{5H} \otimes \ket{0}\bra{0}_{5V} \otimes \ket{0}\bra{0}_{6H} \otimes \ket{1}\bra{1}_{6V} \otimes \mathbb{1}_B$
results in the bit-flipped hybrid states $\ket{V}_A \ket{\alpha_f}_B + e^{i \varphi} \ket{H}_A \ket{-\alpha_f}_B$. It can be converted to the target state by performing a simple bit-flip operation on mode $A$ or a $\pi$-phase shift on mode $B$. The total success probability is therefore $P_{\rm tot}^\varphi=2P^\varphi$. We can also change the relative phase of $|\chi\rangle_{12}$ in order to change the relative phase $\varphi$ of the generated hybrid state.

\section{Experimental Considerations}

\subsection{Detection inefficiency and vacuum mixtures}

\begin{figure}[t]
\includegraphics[width=0.85\linewidth]{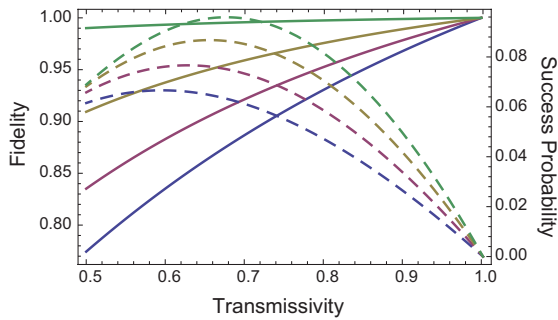}
\caption{(Color online) Fidelity (solid curves) and total success probability (dashed curves) of the hybrid entangled state $\ket{\Psi_\pi(\alpha_f)}_{AB}$ 
 for the beam-splitter transmissivity $t$. The amplitude of the target hybrid state is assumed to be $\alpha_f =1$, and
four cases are plotted with detection efficiencies $\eta = 0.7$, $0.8$, $0.9$, and $0.99$ (starting from the bottom).}
\label{TransmissivityFP}
\end{figure}
\begin{figure}[b]
\includegraphics[width=1.0\linewidth]{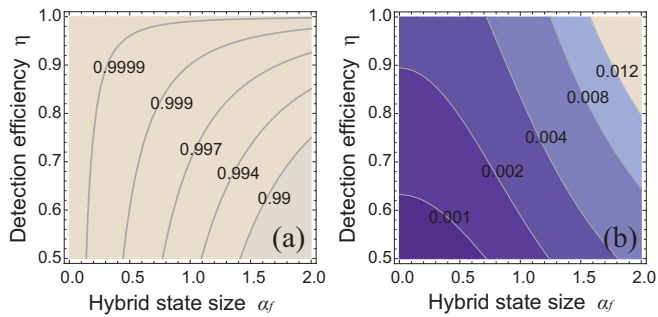}
\caption{(Color online) (a) Fidelity and (b) total success probability for state $\ket{\Psi_\pi(\alpha_f)}_{AB}$  in terms of its amplitude ($\alpha_f$) and detection efficiency ($\eta$). The transmissivity of the beam splitter is assumed to be $t=0.99$.}
\label{HybridFP}
\end{figure}

We need to consider effects of imperfect photodetectors that may lower the fidelity between the generated hybrid state and the ideal one.
An imperfect photodetector with quantum efficiency $\eta$ can be expressed as a positive operator-valued measurement 

\begin{equation}
\hat{E}^{(n)}_\eta = \sum_{m=0}^{\infty} \binom{n+m}{m} \eta^n (1-\eta)^m \ket{n+m} \bra{n+m}
\end{equation}
in the photon number basis.
The total measurement operator for our scheme described in Fig.~1 then becomes 
\begin{equation}
\Pi_\eta = \mathbb{1}_A \otimes \hat{E}^{(0)}_{\eta, 5H} \otimes \hat{E}^{(1)}_{\eta,5V} \otimes \hat{E}^{(1)}_{\eta,6H} \otimes \hat{E}^{(0)}_{\eta,6V} \otimes \mathbb{1}_B,
\label{M1}
\end{equation}
and the heralded state is given by
\begin{equation}
\rho_\eta = \frac{{\rm Tr}_{56} \left[ {\Pi_\eta \ket{\psi_\varphi}\bra{\psi_\varphi}} \right]}{\bra{\psi_\varphi} \Pi_\eta \ket{\psi_\varphi}}.
\end{equation}
In the case of imperfect detection, the fidelity and the success probability can be calculated as 
\begin{equation}
{\cal F}_{\eta}^\varphi = _{AB}\bra{\Psi_\varphi} \rho_\eta \ket{\Psi_\varphi}_{AB}= \frac{1}{2} \left( 1+ e^{-2  (1-\eta)  ( \frac{1}{t} -1 ) \alpha_f^2} \right)
\label{eq:fidelity-imperf}
\end{equation}
and
\begin{equation}
P_{\eta, {\rm tot}}^\varphi = 2 \bra{\psi_\varphi} \Pi_\eta \ket{\psi_\varphi} 
= 2 N_\varphi^2 \eta^2 (\frac{1}{t} - 1)  \alpha_f^2 e^{-2 \eta  (\frac{1}{t}-1 ) \alpha_f^2 },
\end{equation}
respectively.
The fidelity and the success probability of the heralded state depend on $\eta$, $\alpha_f$\, and $t$.
We emphasize that as shown in Eq.~(\ref{eq:fidelity-imperf}),
even if the detection efficiency $\eta$ is limited, the hybrid state can be generated with an arbitrarily high fidelity by taking $t \rightarrow 1$. The cost to obtain a high fidelity is to tolerate a low success probability which becomes zero as the fidelity reaches unity.
Figures \ref{TransmissivityFP} and \ref{HybridFP} show the fidelity and the success probability by changing various parameters.

In a real experiment, the polarization entangled photon pair $\ket{\chi}_{12}$ used for our scheme
may be mixed with the vacuum state $\ket{0}_{12}$ for modes 1 and 2.
The effective form of 
such a mixed state is 
\begin{equation}
\rho_{\chi} =  z \big(\ket{\chi}\bra{\chi}\big)_{12}+(1-z)\big (\ket{0}\bra{0}\big)_{12} ,
\end{equation}
where $0 < z \le 1$.
Remarkably, the vacuum component can be filtered by the conditioning measurement $\Pi$.
When states $\ket{0}_{12} \otimes \ket{{\rm SCS}_\varphi(\alpha_i)}_3$ are initially prepared, the states for modes 5 and 6 will become $\ket{\sqrt{2r}\alpha_i}_5 \ket{0}_6$ or $\ket{0}_5 \ket{\sqrt{2r}\alpha_i}_6$ before the heralding measurement [see Eq.~(\ref{pre-state})], and one of the modes will not contain any photons.
Therefore, there is no chance to get the successful measurement event (i.e., single-photon measurement on both modes 5 and 6).
Meanwhile, the success probability decreases by factor $z$ as the procedure starting with the vacuum state always fails.

\subsection{Use of approximate resource states}

\begin{figure}[t]
\includegraphics[width=1\linewidth]{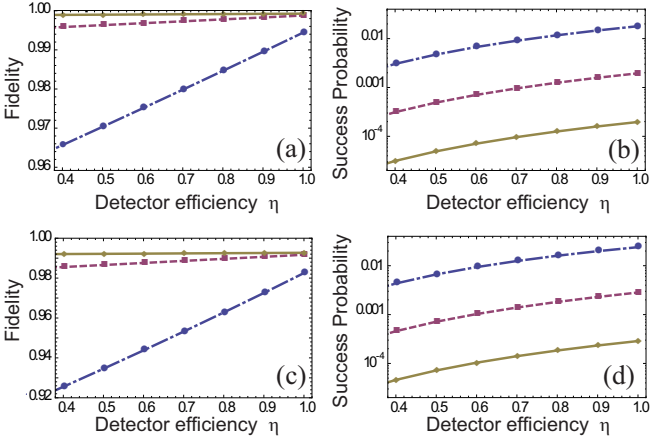}
\caption{(Color online) (a), (c) Fidelity and (b), (d) total success probability for state $\ket{\Psi_\pi(\alpha_f)}_{AB}$ using
photon-subtracted squeezed states as approximate SCSs. The squeezing parameters used to obtain the photon-subtracted squeezed states are $s=0.161$ (upper figures) and $s=0.313$ (lower figures). The transmissivity is $t=0.9$ (dot-dashed lines), $t=0.99$ (dashed lines), and $t=0.999$ (solid lines), respectively.
The vacuum portion of the polarization entangled pair is assumed to be $1-z=0.5$.}
\label{SqueezeFP}
\end{figure}

The SCSs required as resources for our scheme have been experimentally demonstrated while their fidelities and sizes are more or less limited \cite{Ourjoumtsev2006,Neergaard-Nielsen2006,Ourjoumtsev2007,Takahashi2008,Gerritt2010}.
As an example, 
it has been shown that a photon-subtracted squeezed state (or equivalently, a squeezed single photon \cite{note1})
well approximates an ideal SCS, $\ket{{\rm SCS_\pi}(\alpha)} \propto \ket{\alpha} - \ket{-\alpha} $, for relatively small values of  $\alpha$  \cite{Lund2004,JeongLundRalph2005}, and its experimental demonstrations have been reported \cite{Ourjoumtsev2006,Neergaard-Nielsen2006,Takahashi2008,Gerritt2010}.
A squeezed single-photon state in the Fock basis is
\begin{equation}
\hat{S}(s) \ket{1} = \sum_{n=0}^{\infty} \frac{(\tanh{s})^n}{(\cosh{s})^{3/2}} \frac{\sqrt{(2n+1)!}}{2^n n!} \ket{2n+1},
\label{Squeezed-State}
\end{equation}
where $\hat{S}(s) = e^{-(s/2)(\hat{a}^2 - \hat{a}^{\dagger 2})}$ and $s$ is the squeezing parameter.
Its fidelity to an ideal state $\ket{{\rm SCS_\pi}(\alpha)}$ is
\begin{equation}
{\cal F}(\alpha,s) = | \bra{ {\rm SCS}_\pi(\alpha)} \hat{S}(s) \ket{1}|^2 = \frac{2 \alpha^2 e^{\alpha^2 (\tanh{s} -1) } }{(\cosh{s})^3 (1-e^{-2\alpha^2})}.
\end{equation}
For example, squeezing parameters $s=0.161$ and $0.313$ approximate $\ket{{\rm SCS_\pi}(\alpha_i)}$ with amplitudes $\alpha_i = 0.7$ and $ 1 $ with fidelities
${\cal F} = 0.9998$ and $0.997$, respectively
\cite{Lund2004}. We choose these two values for our investigation.

We note that for a small squeezing parameter $s$, it is sufficient to reduce the state (\ref{Squeezed-State})
in the number basis with an appropriate cutoff number, $n_{\rm cut}$, for our numerical calculations.
For example,
the amplitude ratio of $n=7$ to $n=0$ of state (\ref{Squeezed-State}) is less than $0.0005$ for $s= 0.313$ (and even smaller for $s= 0.161$), thus we take the cut-off number $n_{\rm cut}=7$, where the actual photon number cutoff is $2n_{\rm cut} +1 = 15$ from Eq.~(\ref{Squeezed-State}).
We can also model the beam-splitter of transmissivity $t$ ($r = 1-t$) in the photon number basis, which transforms incoming modes $i$ and $j$ into outgoing modes $i'$ and $j'$ as
\begin{equation}
\ket{n}_i \ket{m}_j  \rightarrow \sum_p^n \sum_q^m B_{pq} \ket{p+m-q}_{i'} \ket{n-p+q}_{j'},
\end{equation}
where $B_{pq} = \large[ \binom{n}{p} \binom{m}{q} t^{p+q} r^{n+m-p-q} \large]^{1/2} (-1)^{n-p}$.
Numerical calculations using $n_{\rm cut}$ and the beam splitter model in the photon number basis are applied in order to calculate the fidelity and the success probability with approximate resource states.
Figure \ref{SqueezeFP} shows that the squeezing parameter of $s= 0.161$ $(s=0.313)$ and the vacuum portion of 
$z = 0.5$ result in the fidelity of the heralded hybrid entanglement with fidelity
${\cal F} > 0.996$ $({\cal F} > 0.986)$ and amplitude $\alpha_f \approx 0.7$ $(\alpha_f \approx1.0)$ by taking transmissivity $t \geq 0.99$ and assuming realistic detector efficiency $\eta \geq 0.4$.
We emphasize that the two chosen amplitudes here, $\alpha_f \approx0.7$ and 
$\alpha_f \approx1.0$, for hybrid entanglement were suggested as the best values for a loophole-free Bell test \cite{Kwon2013} and for the hybrid-qubit quantum computation \cite{SWLee2013}, respectively.
The success probability of the conditioning measurement
with $t=0.99$ 
varies from
$P_{\rm tot} \approx 10^{-4}$
to $P_{\rm tot} \approx 10^{-3}$
by increasing the detection efficiency $\eta$ from $0.4$ to $1$.

In order to investigate a degree of entanglement for the heralded hybrid states, we evaluate negativity of the partial transpose \cite{Peres,Horo,Lee2000},
$E (\rho) = ||\rho^{T_A}|| -1 = -2 
 \sum_i \lambda_i^-$,
where $\rho^{T_A}$ is the partial transpose of $\rho$ and $\lambda_i^-$ are its negative eigenvalues.
The degree $E(\rho)$ 
ranges from $0$ to $1$, while an ideal hybrid state of $\alpha\gg1$ results in $E(\rho) \approx 1$.
The degrees of entanglement are  $E(\rho)=0.922$ $(E(\rho)=0.982)$ for squeezing parameters $s=0.161$ $(s=0.313)$
by taking $t=0.99$, $z=0.5$, and $\eta=0.7$.
The entanglement degrees can be compared with  those of the ideal hybrid states with $\alpha_f = 0.7$ and $\alpha_f = 1.0$, i.e., $E(\rho) = 0.927$ and $E(\rho) =0.991$, respectively.

\subsection{Imperfect on-off detectors and SPDC sources}

An on-off photodetector (e.g., avalanche photodiode) typically used in a laboratory does not distinguish between a single photon and two or more photons.
Furthermore, a realistic polarized photon pair generated by spontaneous parametric down conversion (SPDC) contains undesired vacuum and higher order terms in addition to state $\ket{\chi}$.

On-off photodetection changes the conditioning measurement of Eq.~(\ref{M1}) to
\begin{equation}
\Pi^{\rm on-off}_\eta = \mathbb{1}_A \otimes \hat{E}^{(0)}_{\eta, 5H} \otimes \hat{E}^{(\rm click)}_{\eta,5V} \otimes \hat{E}^{(\rm click)}_{\eta,6H} \otimes \hat{E}^{(0)}_{\eta,6V} \otimes \mathbb{1}_B,
\end{equation}
where $\hat{E}^{(\rm click)} = \mathbb{1} - \hat{E}^{(0)} = \sum_{m=0}^{\infty} [ 1 -  (1-\eta)^m ] \ket{m} \bra{m}$.
The polarization entangled state created by SPDC can be represented by
$\ket{{\rm SPDC}_\chi} = \exp(\xi\hat{K}_+ + \xi^* \hat{K}_-) \ket{0}_{12}$,
where $\hat{K}_+ = \hat{a}^\dagger_{1H} \hat{a}^\dagger_{2V} + \hat{a}^\dagger_{1V} \hat{a}^\dagger_{2H}$ and $\hat{K}_-=\hat{K}_+^\dagger$
with the squeezing parameter $\xi$.
The state can be simplified as 
\begin{equation}
\ket{{\rm SPDC}_\chi} 
= \sqrt{1-\lambda^2} \sum_{n=0}^{\infty} \lambda^n \ket{\Phi_n}_{12},
\end{equation}
where $\lambda=\tanh{|\xi|}$ is the interaction strength and $\ket{\Phi_n}_{12} = (n+1)^{-1/2} \sum_{m=0}^n \ket{m}_{1H}\ket{n-m}_{1V}\ket{n-m}_{2H}\ket{m}_{2V}$  \cite{Kok2010}.
In this case, the probability ratio for $\ket{\Phi_n}$ has an order of ${\cal O}(\lambda^{2n})$. 
Note that $\ket{\Phi_0}$ is the vacuum state and $\ket{\Phi_1} = \ket{\chi}$.
The total success probability of the final heralding measurement using the SPDC source then becomes
\begin{equation}
P_{\rm tot} = (1-\lambda^2) \left[ P_{\rm vac} + \lambda^2 P_{\ket{\chi}}  + \lambda^4 P_{\ket{\Phi_2}} + {\cal O}(\lambda^6) \right],
\end{equation}
where $P_{\rm vac}$,  $P_{\ket{\chi}}$, and $P_{\ket{\Phi_2}}$ are success probabilities when the input state was the vacuum, $\ket{\chi}$, and $\ket{\Phi_2}$, respectively. 
Generally, $\lambda$ in the SPDC source has a small value so that higher order terms can be neglected. We shall ignore ${\cal O}(\lambda^6)$ in the following calculations.

\begin{figure}[t]
\includegraphics[width=.7\linewidth]{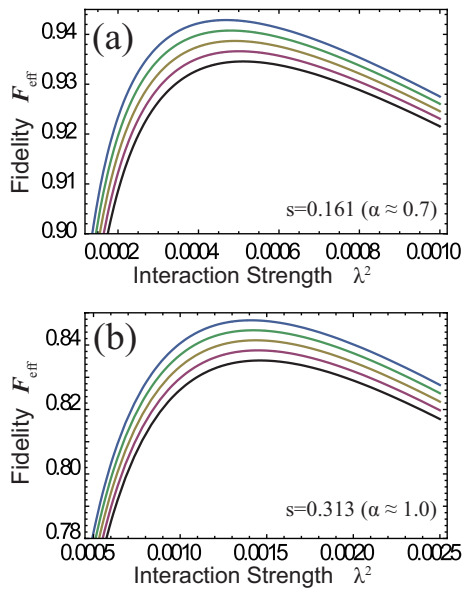}
\caption{(Color online) Expected fidelity of the generated hybrid entanglement when a squeezed single-photon state and a SPDC source with interaction strength $\lambda$ are applied to the scheme using inefficient on-off detectors.
The squeezing parameters are (a) $s=0.161$ and (b) $s=0.313$ while the beam-splitter transmissivity is $t=0.99$ for both cases.
The efficiencies of the on-off detectors are $\eta =0.9$, $0.7$, $0.5$, $0.3$, and $0.1$ starting from the top.}
\label{SPDC}
\end{figure}

The input state of $\ket{\chi}$ is the only desired state for generating the hybrid entanglement and apparently successful heralding measurements of all the other input states will degrade the fidelity of the generated state. 
The fidelity ${\cal F}_{\rm eff}$ of the finally generated state under these more realistic conditions is
\begin{equation}
\begin{aligned}
{\cal F}_{\rm eff} &= \frac{(1-\lambda^2) \lambda^2 P_{\ket{\chi}}} {P_{\rm tot}} {\cal F} \\
& \approx \frac{ P_{\ket{\chi}} } { \lambda^{-2} P_{\rm vac} +  P_{\ket{\chi}}  + \lambda^2 P_{\ket{\Phi_2}}} {\cal F}.
\end{aligned}
\end{equation}
We calculate  $P_{\rm vac}$,  $P_{\ket{\chi}}$ and $P_{\ket{\Phi_2}}$ using the numerical method in the number basis as explained in the previous section.
We  plot the final fidelities ${\cal F}_{\rm eff}$ for several choices of on-off detection efficiencies $\eta$ and the beam-splitter transmissivity $t=0.99$ in Fig.~\ref{SPDC}.
Remarkably, the fidelities are insensitive to inefficiency $\eta$ of the on-off detectors even though it reduces the success probabilities of the scheme. The fidelities are reasonably high for large regions of experimentally relevant values of the interaction strength $\lambda$.
For example, we can obtain the hybrid state of $\alpha_f = 0.7$  and ${\cal F}_{\rm eff} \approx 0.939$ using the SPDC source of $\lambda=2.2 \times10^{-2}$ and a squeezed single-photon state of $s=0.161$ with on-off detectors of 50\% efficiency, while the success probability is reduced to $P_{\rm tot}=5.1\times 10^{-7}$.
As another example, the hybrid state of $\alpha_f = 1$ and ${\cal F}_{\rm eff} \approx 0.842$ can be generated
using the SPDC source of $\lambda=3.8 \times10^{-2}$ and a squeezed single-photon state of $s=0.313$ with
the on-off detectors of 50\% efficiency while the success probability is $P_{\rm tot}=2.4\times10^{-6}$.
Figure~\ref{SPDC} shows that  the fidelities are still reasonably high even when the detection efficiency is as low as 10\%.
We also note that dark counts during the heralding detection process may be another factor to degrade the final fidelity, and photodetectors with ultralow dark count  rates compared to quantum efficiency \cite{Akiba2009, Zhang2011,Schuck2013,ultra, Review2012} may be used for high fidelities.
On the other hand, we expect that the effects of dark counts may be limited at a reasonable level using current technology as done for this type of experiment \cite{Jeong2014,Morin2014}.

\section{Remarks}

We have suggested a scheme to generate hybrid entanglement between a single photon qubit and a coherent state qubit.
Unlike previous proposals \cite{Andersen2013,Jeong2014,Morin2014},
our scheme enables one to generate the exact form of hybrid entanglement, without approximation, required for resource-efficient optical hybrid quantum computation~\cite{SWLee2013} and loophole-free Bell inequality tests \cite{Kwon2013}.
The required resources are
an SCS, an entangled photon pair, the displacement operation, four photodetectors, and beam splitters. 
Even when photodetectors with limited efficiencies are used, hybrid entanglement
with an arbitrarily high fidelity can be generated at the price of a lower success probability.
We have also analyzed fidelities of the generated states when a SPDC source, an approximate SCS, and  on-off detectors with low efficiencies are used for the scheme.
Even under these realistic assumptions, hybrid entanglement with high fidelities may be obtained.
According to our analysis, experimental implementation of our scheme
seems feasible using current technology despite some expected experimental imperfections.

\section*{acknowledgment}
This work was supported by the National Research Foundation of Korea (NRF) through a grant funded by the Korea government (MSIP) (Grant No. 2010-0018295).
H. K. was supported by the Global Ph.D. Fellowship Program through the NRF funded by the Ministry of Education (Grant No. 2012-003435).

%\newpage

\end{document}